# Massless states and negative mass states of the coupled electron-positron system with completely symmetric representation of the particles

A.I. Agafonov

National Research Centre "Kurchatov Institute", Moscow 123182, Russia

E-mail: Agafonov_AIV@nrcki.ru

We argue that the free electron and positron can be considered as different, independent particles, each of which is characterized by the complete set of the Dirac plane waves. This completely symmetric representation of the particles makes it necessary to choose another solution of the Dirac equation for the free particle propagator as compared to that currently used in QED. The Bethe-Salpeter equation is studied in the ladder approximation with using these free propagators.

A new branch of electron–positron bound states which represent the massless composite bosons, have been found for the actual coupling equal to the fine structure constant. We have obtained that: 1) the massless boson states have the normalized complex wave functions; 2) the average distance between the electron and positron diverges as the boson kinetic energy goes to zero; 3) the spatial contraction of the wave function of the transverse motion of strongly coupled electron-positron pair is continuously occurred with increasing the boson kinetic energy. Unlike the usual annihilation process in which nothing remains from the electron and positron, a similar annihilation-like process in which the reaction products are two or three gamma quanta and the massless boson, is predicted.

In this symmetric representation one could expect states, which have a certain symmetry relative to Ps states, but have negative masses. For these states the equal-time bound-state equation was derived neglecting the interaction retardation and interaction through the vector potential. It turned out that the wave functions of the negative mass boson states are not normalized. Beyond these assumptions, the existence of these negative mass states remains unclear.

## 1. Introduction

As is well known [1,2,3], the set of the Dirac plane waves, including both positive and negative energy states, forms a complete orthonormal system. The general solution of the Dirac equation for the free fermion is written as a superposition of all the plane waves. However, Feynman insisted on [4] that there are only the states with the positive energies. This leads to the conclusion that there must be antiparticles, and the hole theory of antiparticles has been developed [5].

In the standard model of the electron-positron field the initially complete basis of the Dirac plane waves is divided into two parts: the states with positive energies are accepted as the electron states, and the states with negative energies are declared the states of the positrons which are recognized as particles traveling backwards in time. The filled states with negative energies are as a rule not considered, and the operator of charge conjugation, which converts the particle into antiparticle and vice versa, are introduced. Then, using the Bogoliubov-like transformation, the description of the electron-positron field can be mathematically made symmetric. However, the base of the standard model is laid initial physical asymmetry due to the introduction of the primary particles which are usually treated as the electrons, and the positrons are considered as the electronic holes in the states of the filled lower continuum [6,7].

Note that in many situations, the filled electronic states with negative energies cannot be ignored. So, the filled lower continuum is important in the analysis of the electronic structure in super-heavy nuclei, for which at the certain charge of the nucleus the electron lower level $1S_{1/2}$ merges with the bottom of the lower continuum [8,9]. A similar situation arises when discussing the value of the cosmological constant. Apart from the positive contribution from the zero-point energy of boson quantized fields, another energy source is derived from the Dirac theory of the electron because the filled levels leads inevitably to negative contribution to the vacuum energy [7]. By studying the radiation scattering by free electrons, it was concluded that radiation-induced electron quantum jumps in the intermediate states of negative energies are crucial for the scattering [10].

There is no reason to doubt that the complete spectrum of states for any system of interacting particles can be deduced only when the full basis of states is taken into account for each particle of the system. The above two part division of the complete plane-wave basis leads to the following fact: neither electron states nor positron states form the complete system of the wave functions. Therefore, there is no reason to believe that all the states of the coupled electron-positron system will be obtained and examined in the standard approach.

The latter statement can be clearly argued on example of the well-known non-relativistic problem of impurity states in semiconductors, as shown in Section 2. Here the complete system of single-electron states includes all states in the allowed bands of the semiconductor. Considering the states of only one band, for example, the conduction band or the valence band, and ignoring all the others, this task is reduced to the so-called effective Hamiltonian in the single-band approximation. The eigenfunctions of this Hamiltonian correspond to the shallow energy levels in the energy gap near the conduction band bottom or valence band top, respectively. Accounting for the complete basis of one-electron states changes substantially the picture of the impurity state spectrum [11], because, apart from energy corrections for shallow states, localized strongly-coupled states can appear deep in the semiconductor forbidden gap.

Following the symmetry of the laws of nature with respect to the two signs of the electric charge, we have to assume that the electron and positron are independent and different ordinary particles, each of which can be described by the Dirac equation. Then both the electron and the positron should be characterized by the complete orthonormal system of the Dirac plane waves, including both positive and negative energy states. This absolutely symmetric representation of the particles makes it necessary to choose another solution of the Dirac equation for the free particle propagator as compared to that currently used in QED, as discussed in Section 3. The Bethe-Salpeter equation in the ladder approximation with these free propagators is the basis for further studies in this work.

Firstly, we study the states for which the characteristic velocity of particle of the bound pair is much smaller than the speed of light in vacuum, as it is for the positronium. It is known that the interaction retardation due to the finite speed of light, and the contribution to the interaction through the vector potential are only resulted in the relativistic corrections to the energy spectrum of positronium. In this article we are not interested in these corrections, and the important issue for us is the complete picture of the spectrum of bound states. Neglecting the above two effects, the equal-time bound state equation is derived in Section 4.

This equation is studied in Section 5. Besides the positronium states the energy levels of which lie just below the energy $2m$ ($m$ is the electron mass), due to this symmetrical representation of the particles it would be possible to analyze the negative mass boson states formed by the coupled electron-positron system. However, these states are not normalized to unity. This equation contains also the solution with energy $E = 0$ that is the only dedicated energy level due to the symmetry of the problem considered. However, the approaches used in the derivation of this equation, does not allow us to affirm that this strongly-coupled state, in which the mass of the composite boson completely disappear because of the high binding energy equal to $-2m$, does exist.

In Section 6, the Bethe-Salpeter equation for the electron-positron system is studied in the ladder approximation with using these free propagators, and for the actual coupling equal to the fine structure constant. We search for the solution of the Bethe-Salpeter equation in the form of a stationary wave with the phase velocity equal to the speed of light in vacuum. This wave is an infinite line string, the transverse size of which is determined by the two dimensional wave function of the transverse motion of the coupled pair. The equation for the transverse wave function is derived with account for both the interaction retardation due to the finite speed of light, and the contribution to the interaction through the vector potential. It turned out that these massless boson states are formed only when the helicities of the electron and positron are opposite, and the transverse wave function satisfy the homogeneous Fredholm integral equation of the second kind with non-Fredholm kernel. Results of numerical studies of this equation are reported.

Discussions of the results obtained, and conclusion are presented in Section 7. The very important process is the electron and positron annihilation accompanied by photons production. The existence of the composite massless boson must lead to some process which is very similar to the annihilation in the standard model, but has the essential distinctive features due to the fact that the products of this reaction include, in addition to photons, the massless boson. So, for the singlet electron–positron pair with zero total momentum the annihilation radiation spectra must have finite both angular and momentum correlations.

Natural units ($\hbar = c = 1$) will be used throughout.

## 2. Strongly bound states in the non-relativistic problem

Consider a two-band semiconductor, the eigenfunctions $\psi_{n\mathbf{q}}(\mathbf{r})$ ($n=1,2$) and eigenenergies $\varepsilon_n(\mathbf{q})$ of which are given by the Schrödinger equation $H\psi_{n\mathbf{q}}(\mathbf{r}) = \varepsilon_n(\mathbf{q})\psi_{n\mathbf{q}}(\mathbf{r})$. Here $H$ is the Hamiltonian of the semiconductor. Let the upper band ($n=1$) is empty and the lower band ($n=2$) is completely filled with electrons. Introduce $V(\mathbf{r})$ - the potential created by an impurity atom inserted into the semiconductor.

Denote $\psi(1)$ as an impurity wave function, where the symbol 1 implies the set of space-time variables $\mathbf{r}_1, t_1$. Then the wave function $\psi(2)$ at subsequent times $t_2 > t_1$ in the point $\mathbf{r}_2$ is determined by the homogeneous integral equation:

$$\psi(2) = -i\int d\mathbf{r}_1 \int dt_1 K_0(2,1) V(1) \psi(1), \qquad (1)$$

where $K_0(2,1)$ is the retarded Green's function which satisfies the equation:

$$(i\partial/\partial t_2 - H_2)K_0(2,1) = i\delta(2,1), \qquad (2)$$

provided that $K_0(2,1) = 0$ at $t_2 < t_1$. Here the subscript 2 on the operators in the left side of (2) means that the operators act on the variables 2 of $K_0(2,1)$, and $\delta(2,1) = \delta(\mathbf{r}_2 - \mathbf{r}_1)\delta(t_2 - t_1)$.

From (2), we obtain:

$$K_0(2,1) = \sum_{\mathbf{q}n} \psi_{n\mathbf{q}}(\mathbf{r}_2)\psi^*_{n\mathbf{q}}(\mathbf{r}_1) \exp(-i\varepsilon_n(\mathbf{q})(t_2 - t_1))\theta(t_2 - t_1). \qquad (3)$$

We represent the impurity wave function in the form $\psi(1) = \psi(\mathbf{r}_1)\exp(-iEt_1)$. Substituting (3) into (1), and integrating over $t_1$, we obtain the following integral equation for bound states:

$$\psi(\mathbf{r}_2) = \int d\mathbf{r}_1 \sum_{\mathbf{q}} \left\{ \frac{\psi_{1\mathbf{q}}(\mathbf{r}_2)\psi_{1\mathbf{q}}^*(\mathbf{r}_1)}{E - \varepsilon_1(\mathbf{q})} + \frac{\psi_{2\mathbf{q}}(\mathbf{r}_2)\psi_{2\mathbf{q}}^*(\mathbf{r}_1)}{E - \varepsilon_2(\mathbf{q})} \right\} V(\mathbf{r}_1)\psi(\mathbf{r}_1). \qquad (4)$$

If one neglects the second term in the curly brackets in the right-hand side of (4), that corresponds to the neglect of states of the lower band 2, this equation is reduced to the effective single-band Hamiltonian:

$$(\varepsilon_1(-i\nabla_{\mathbf{r}}) + V(\mathbf{r}))\psi = E\psi.$$

It is well known that the eigenfunctions of this single-band Hamiltonian correspond to shallow energy levels near the bottom of the band 1.

However, when taking into account the valence band 2 in Eq. 4 the situation changes significantly. Then, the integral equation (4) can be reduced to the following differential equation [11]:

$$(E - \varepsilon_1(-i\nabla_{\mathbf{r}}))(E - \varepsilon_2(-i\nabla_{\mathbf{r}}))\psi = (2E - \varepsilon_1(-i\nabla_{\mathbf{r}}) - \varepsilon_2(-i\nabla_{\mathbf{r}}))V(\mathbf{r})\psi. \qquad (5)$$

Eq. (5), along with the shallow states, has a solution for a strongly coupled bound state. It was found [11] that for the parabolic dispersion law for both the bands with the equal effective masses this impurity energy level is exactly in the middle of the semiconductor energy gap.

Note that there are other known methods for calculation of strongly coupled impurity levels (see [12,13] and references therein). Common to them is that the deep levels that are actually observed in experiments appear in multiband models of semiconductors.

Thus, on the example of the impurity semiconductor it was shown that for interacting particle systems the full spectrum of bound states can be found only when the full basis of states is taken into account for each particle of the system.

## 3. The free fermion propagator

At present in QED the free fermion propagator for the Dirac equation is written as [1,2,3]:

$$K_+(2,1) = \sum_{\mathbf{p}} \psi_p(2)\overline{\psi}_p(1)\theta(t_2 - t_1) - \sum_{\mathbf{p}} \psi_{-p}(2)\overline{\psi}_{-p}(1)\theta(t_1 - t_2), \qquad (6)$$

where $\psi_{\pm p}$ is the Dirac plane wave representing the state of the free particle with energy $\pm\varepsilon_{\mathbf{p}}$,

respectively, and $\bar{\psi}_p = \psi_p^* \beta$ denotes the Dirac conjugation. In Eq. (6) the contribution to $K_+(2,1)$ at $t_2 > t_1$ is due to the electron terms, and at $t_2 < t_1$ - the positron terms.

The Bethe-Salpeter equation [14,15] with the propagator (6) has been studied in works, as a rule, in the ladder approximation. The results obtained can be summarized as follows. Firstly, the weakly bound states of positronium have been found with certain relativistic corrections. Secondly, after the work [16], considerable interest is the problem of strongly coupled states for fermion-antifermion systems. In the most commonly used approach to the problem the Bethe-Salpeter equation is regarded as eigenvalues task for the coupling constant [17-25]. That is, an eigenvalue is considered as the necessary strength of the attractive potential to make a massless bound state.

Thirdly, the Feynman theory leads to a clearer picture of the annihilation of electron-positron pairs [4]. In principal this process can be studied the Bethe-Salpeter equation in the ladder approximation with account for the "annihilation" interaction which can be treated as a perturbation [26]. Therefore, any theory of the electron-positron field must include a process that, at least, looks the annihilation-like process.

The Dirac equation, as well as any differential equation, has several solutions for Green's function [27]. For example, the retarded Green function of the following form was discussed in [5]:

$$K_-(2,1) = \sum_{\mathbf{p}} \left( \psi_p(2) \bar{\psi}_p(1) + \psi_{-p}(2) \bar{\psi}_{-p}(1) \right) \theta(t_2 - t_1). \qquad (7)$$

Comparing (6) and (7), one can conclude that unlike the propagator (7) the negative energy states are actually excluded from consideration in (6). Obviously that in this case neither electron states nor positron states form a complete orthonormal system of the wave functions. It can be argued that this approach is equivalent to the single-band approximation in the sense, as discussed in the previous section 2. Hence, there is no reason to believe that the whole spectrum of bound states of the electron-positron system will be obtained with using of (6).

Considering (7) as the electron propagator, we can see that this function for the electron is taken into account the whole spectrum of the Dirac plane waves. There is no doubt that the positron can be described by the Dirac equation as well. Then we have the only opportunity, which is to assume that the electron and positron are independent and different ordinary particles.

In this approach, in the vacuum state the upper continuum is empty and the lower continuum is completely filled for each of these particles. Then this vacuum state is charge neutral.

Similarly to (7), for the positron propagator we would have:

$$K_+(4,3) = \sum_{\mathbf{p}} \left( \varphi_p(4)\overline{\varphi}_p(3) + \varphi_{-p}(4)\overline{\varphi}_{-p}(3) \right) \theta(t_4 - t_3). \tag{8}$$

In Eqs (7) and (8) $\psi_{\pm p}$ и $\varphi_{\pm p}$ are the Dirac plane waves for the free electrons and positrons, respectively.

Expressions (7) and (8) can be regarded as a consequence of the symmetric representation of the particles. It is important that regardless of whether the electron or positron in the states of the upper or the lower continuum, the interaction between the particles is attractive. In the ladder approximation their reaction can be presented by the retarded interaction function [5,8]:

$$G^{(1)}(3,4;5,6) = -e^2(1-\boldsymbol{\alpha}_-\boldsymbol{\alpha}_+)\delta_+(s_{56}^2)\delta(3,5)\delta(4,6). \tag{9}$$

Here $\boldsymbol{\alpha}_\pm = \begin{pmatrix} 0 & \boldsymbol{\sigma}_\pm \\ \boldsymbol{\sigma}_\pm & 0 \end{pmatrix}$ are the velocity operators for the electron (-) and positron (+), and $\boldsymbol{\sigma}_\pm$ are the Pauli matrices for the electron and positron, respectively, and $s_{56}$ is the invariant distance between particles.

Although the propagators (7) and (8) are taken into account the complete orthonormal system of the wave functions for each of the particles, the interaction between them is attractive only if both these particles are in the states with the positive (negative) energies. If one of them is in the states of the upper continuum, and the other - in the states of the lower continuum, the interaction between the particles changes sign, and becomes repulsive. Therefore propagators (7) and (8) are not suitable.

Taking into account the character of the interaction of the particles, the electron retarded Green function should be written as:

$$K_{0-}(2,1) = \sum_{\mathbf{p}} \left( \psi_p(2)\psi_p^+(1) + \psi_{-p}(2)\psi_{-p}^+(1) \right) \theta(t_2 - t_1), \tag{10}$$

and, similarly, for the positron propagator

$$K_{0+}(4,3) = \sum_{\mathbf{p}} \left( \varphi_p(4)\varphi_p^+(3) + \varphi_{-p}(4)\varphi_{-p}^+(3) \right) \theta(t_4 - t_3). \tag{11}$$

Here $\psi_{\pm p}^+$ и $\varphi_{\pm p}^+$ are the Hermitian conjugate matrices with respect to $\psi_{\pm p}$ and $\varphi_{\pm p}$, respectively. The letter are given by [7]: $\psi_p, \varphi_p = u_{p,\pm} e^{-ipx/\hbar}$ and $\psi_{-p}, \varphi_{-p} = u_{-p,\pm} e^{ipx/\hbar}$, where

$$u_{p,\pm} = \frac{1}{\sqrt{2\varepsilon_p}} \begin{pmatrix} \sqrt{\varepsilon_p + mc^2}\, w_\pm \\ \sqrt{\varepsilon_p - mc^2}\, (\mathbf{n}\boldsymbol{\sigma}_\pm) w_\pm \end{pmatrix} \quad u_{-p,\pm} = \frac{1}{\sqrt{2\varepsilon_p}} \begin{pmatrix} \sqrt{\varepsilon_p - mc^2}\, (\mathbf{n}\boldsymbol{\sigma}_\pm) w_\pm' \\ \sqrt{\varepsilon_p + mc^2}\, w_\pm' \end{pmatrix}. \tag{12}$$

Using (9)-(11), in the ladder approximation the bound state Bethe-Salpeter equation for the electron-positron system is:

$$\psi(1,2) = -i \iiiint d\tau_3 d\tau_4 d\tau_5 d\tau_6 K_{0-}(1,3) K_{0+}(2,4) G^{(1)}(3,4;5,6) \psi(5,6), \quad (13)$$

where $d\tau_i = d\mathbf{r}_i dt_i$.

## 4. Quasi non-relativistic limit

Bound states derived from the Bethe-Salpeter equation, have greater reliability if they have a non-relativistic analog. With this motivation we study the states for which the characteristic velocity of particle of the bound pair is much smaller than the speed of light in vacuum, $v/c \propto \alpha \ll 1$, where $\alpha$ is the fine structure constant. Such states are known to be inherent to positronium.

The interaction retardation and the interaction through the vector potential lead only to the relativistic corrections to the energy spectrum of positronium. In this Section they are out of our interests. Neglecting the above two effects, the interaction function in (9) is replaced by:

$$G^{(1)}(3,4;5,6) = V(r_{56}) \delta(t_{56}) \delta(3,5) \delta(4,6). \quad (14)$$

where $V(r_{56}) = -\dfrac{e^2}{r_{56}}$ is the Coulomb interaction. As a result, Eq. (13) is reduced to the equal-time equation:

$$\psi(\mathbf{r}_1, \mathbf{r}_2; t) = -i \int d\mathbf{r}_3 \int d\mathbf{r}_4 \int_{-\infty}^{t} dt_1 K_{0-}(\mathbf{r}_1 - \mathbf{r}_3; t - t_1) K_{0+}(\mathbf{r}_2 - \mathbf{r}_4; t - t_1) V(|\mathbf{r}_3 - \mathbf{r}_4|) \psi(\mathbf{r}_3, \mathbf{r}_4; t_1).$$

The wave function is presented in the form $\psi(\mathbf{r}_1, \mathbf{r}_2; t) = \psi(\mathbf{r}_1, \mathbf{r}_2) \exp(-iEt)$, where $E$ is the energy of the coupled state. Then after the integration over time the above equation takes the form:

$$\psi(\mathbf{r}_1, \mathbf{r}_2) = \int d\mathbf{r}_3 \int d\mathbf{r}_4 \sum_{\mathbf{pq}} F(\mathbf{p},\mathbf{q}) \exp\left(i\mathbf{p}(\mathbf{r}_1 - \mathbf{r}_3) + i\mathbf{q}(\mathbf{r}_2 - \mathbf{r}_4)\right) V(\mathbf{r}_3 - \mathbf{r}_4) \psi(\mathbf{r}_3, \mathbf{r}_4), \quad (15)$$

where, using (12), the function $F$ is:

$$F = \frac{1}{4\varepsilon_{\mathbf{p}}\varepsilon_{\mathbf{q}}} \left[ \frac{(\varepsilon_{\mathbf{p}} + m_-\beta_- + \boldsymbol{\alpha}_-\mathbf{p})(\varepsilon_{\mathbf{q}} + m_+\beta_+ + \boldsymbol{\alpha}_+\mathbf{q})}{E - \varepsilon_{\mathbf{p}} - \varepsilon_{\mathbf{q}}} + \frac{(\varepsilon_{\mathbf{p}} - m_-\beta_- - \boldsymbol{\alpha}_-\mathbf{p})(\varepsilon_{\mathbf{q}} + m_+\beta_+ + \boldsymbol{\alpha}_+\mathbf{q})}{E + \varepsilon_{\mathbf{p}} - \varepsilon_{\mathbf{q}}} + \right.$$

$$\left. \frac{(\varepsilon_{\mathbf{p}} + m_-\beta_- + \boldsymbol{\alpha}_-\mathbf{p})(\varepsilon_{\mathbf{q}} - m_+\beta_+ - \boldsymbol{\alpha}_+\mathbf{q})}{E - \varepsilon_{\mathbf{p}} + \varepsilon_{\mathbf{q}}} + \frac{(\varepsilon_{\mathbf{p}} - m_-\beta_- - \boldsymbol{\alpha}_-\mathbf{p})(\varepsilon_{\mathbf{q}} - m_+\beta_+ - \boldsymbol{\alpha}_+\mathbf{q})}{E + \varepsilon_{\mathbf{p}} + \varepsilon_{\mathbf{q}}} \right]. \quad (16)$$

Here **p** and **q** are the momentum of electron and positron, respectively, the matrices $\beta_-, \alpha_-$ и $\beta_+, \alpha_+$ act on the electron and positron bispinors of the wave function $\psi$. In Eq. (16) the electron and positron operators can be reordered.

In the expression (16) the masses are marked by $\pm$. Of course, the electron mass is equal to the positron one. However now we can examine the limit $m_+ \to \infty$ in Eq. (15) with definition (16). This case corresponds to the motion of the electron in the external field generated by the fixed positively charged center (for example, by proton).

Assuming that $\varepsilon_\mathbf{q} = m_+$, $\mathbf{q} = 0$ and $E - m_+ \ll E + m_+$, the expression (16) is reduced to:

$$F = \frac{1}{4\varepsilon_\mathbf{p}} \left( \frac{(\varepsilon_\mathbf{p} + m_-\beta_- + \boldsymbol{\alpha}_-\mathbf{p})(1 + \beta_+)}{E - \varepsilon_\mathbf{p} - m_+} + \frac{(\varepsilon_\mathbf{p} - m_-\beta_- - \boldsymbol{\alpha}_-\mathbf{p})(1 + \beta_+)}{E + \varepsilon_\mathbf{p} - m_+} \right). \tag{17}$$

Substituting (17) in (15) and replacing $E - m_+ \to E$, after integration over $\mathbf{r}_4$ we obtain:

$$\psi(\mathbf{r}_1) = \int d\mathbf{r}_3 \sum_\mathbf{p} \frac{1}{2\varepsilon_\mathbf{p}} \left( \frac{(\varepsilon_\mathbf{p} + m_-\beta_- + \boldsymbol{\alpha}_-\mathbf{p})}{E - \varepsilon_\mathbf{p}} + \frac{(\varepsilon_\mathbf{p} - m_-\beta_- - \boldsymbol{\alpha}_-\mathbf{p})}{E + \varepsilon_\mathbf{p}} \right) \begin{pmatrix} 1 & 0 \\ 0 & 0 \end{pmatrix}_+ e^{i\mathbf{p}(\mathbf{r}_1-\mathbf{r}_3)} V(\mathbf{r}_3) \psi(\mathbf{r}_3). \tag{18}$$

Here the radius-vectors $\mathbf{r}_1, \mathbf{r}_3$ are counted from the fixed center position (the vector $\mathbf{r}_2$ in Eq. (15)). From (18) we have for the electron wave function:

$$\psi_-(\mathbf{r}_1) = \int d\mathbf{r}_3 \sum_\mathbf{p} \frac{E + m_-\beta_- + \boldsymbol{\alpha}_-\mathbf{p}}{E^2 - \varepsilon_\mathbf{p}^2} e^{i\mathbf{p}(\mathbf{r}_1-\mathbf{r}_3)} V(\mathbf{r}_3) \psi_-(\mathbf{r}_3).$$

This integral equation is corresponded the following differential equation:

$$(E + m_-\beta_- + \boldsymbol{\alpha}_-\mathbf{p})^{-1} \psi_-(\mathbf{r}) = V(\mathbf{r}) \psi_-(\mathbf{r}).$$

Here $(E + m_-\beta_- + \boldsymbol{\alpha}_-\mathbf{p})^{-1}$ means the inverse matrix. The last equation is the Dirac equation, $(m_-\beta_- + \boldsymbol{\alpha}_-\mathbf{p})\psi_-(\mathbf{r}) = (E - V(\mathbf{r}))\psi_-(\mathbf{r})$.

Now we find a differential equation, which corresponds to the integral equation (15) with (16). To this end, the right side of (16) is reduced to the form:

$$F(\mathbf{p},\mathbf{q}) = \frac{1}{\Delta} \Big[ E\left(E^2 - \varepsilon_\mathbf{p}^2 - \varepsilon_\mathbf{q}^2\right) + 2E(\beta_- m + \boldsymbol{\alpha}_-\mathbf{p})(\beta_+ m + \boldsymbol{\alpha}_+\mathbf{q})$$
$$+ \left(E^2 - \varepsilon_\mathbf{p}^2 + \varepsilon_\mathbf{q}^2\right)(\beta_- m + \boldsymbol{\alpha}_-\mathbf{p}) + \left(E^2 + \varepsilon_\mathbf{p}^2 - \varepsilon_\mathbf{q}^2\right)(\beta_+ m + \boldsymbol{\alpha}_+\mathbf{q}) \Big], \tag{19}$$

where $\Delta = E^4 - 2E^2\left(\varepsilon_\mathbf{p}^2 + \varepsilon_\mathbf{q}^2\right) + \left(\varepsilon_\mathbf{p}^2 - \varepsilon_\mathbf{q}^2\right)^2$.

Representing (19) as:

$$F(\mathbf{p},\mathbf{q}) = \frac{1}{\Delta}(A_0 + (a_- + b_-)(a_+ + b_+)), \qquad (20)$$

and equating (19) and (20), we have:

$$A_0 = \frac{\Delta}{2E}, \quad a_- = \sqrt{2E}(\beta_- m + \boldsymbol{\alpha}_- \mathbf{p}), \quad a_+ = \sqrt{2E}(\beta_+ m + \boldsymbol{\alpha}_+ \mathbf{q}),$$

and

$$b_- = \frac{E^2 + \varepsilon_{\mathbf{p}}^2 - \varepsilon_{\mathbf{q}}^2}{\sqrt{2E}}, \quad b_+ = \frac{E^2 - \varepsilon_{\mathbf{p}}^2 + \varepsilon_{\mathbf{q}}^2}{\sqrt{2E}}.$$

Hence, the function (16) is rewritten as:

$$F(\mathbf{p},\mathbf{q}) = \frac{1}{2E} + \frac{2E}{\Delta}\left(\beta_- m + \boldsymbol{\alpha}_- \mathbf{p} + \frac{E^2 + \varepsilon_{\mathbf{p}}^2 - \varepsilon_{\mathbf{q}}^2}{2E}\right)\left(\beta_+ m + \boldsymbol{\alpha}_+ \mathbf{q} + \frac{E^2 - \varepsilon_{\mathbf{p}}^2 + \varepsilon_{\mathbf{q}}^2}{2E}\right), \qquad (21)$$

Substituting (21) into (15), we obtain:

$$\left(1 - \frac{V(\mathbf{r}_1 - \mathbf{r}_2)}{2E}\right)\psi(\mathbf{r}_1, \mathbf{r}_2) = \int d\mathbf{r}_3 \int d\mathbf{r}_4 \sum_{\mathbf{pq}} \frac{2E}{\Delta}\left(m\beta_- + \boldsymbol{\alpha}_- \mathbf{p} + \frac{E^2 + \varepsilon_{\mathbf{p}}^2 - \varepsilon_{\mathbf{q}}^2}{2E}\right)\left(m\beta_+ + \boldsymbol{\alpha}_+ \mathbf{q} + \frac{E^2 - \varepsilon_{\mathbf{p}}^2 + \varepsilon_{\mathbf{q}}^2}{2E}\right)$$

$$\exp(i\mathbf{p}(\mathbf{r}_1 - \mathbf{r}_3) + i\mathbf{q}(\mathbf{r}_2 - \mathbf{r}_4))V(\mathbf{r}_3 - \mathbf{r}_4)\psi(\mathbf{r}_3, \mathbf{r}_4). \qquad (22)$$

The integral equation (22) corresponds to the following differential equation:

$$\left(E^4 - 2E^2(\varepsilon_{\mathbf{p}}^2 + \varepsilon_{\mathbf{q}}^2) + (\varepsilon_{\mathbf{p}}^2 - \varepsilon_{\mathbf{q}}^2)^2\right)\left(1 - \frac{V(\mathbf{r}_1 - \mathbf{r}_2)}{2E}\right)\psi(\mathbf{r}_1, \mathbf{r}_2) =$$

$$2E\left(m\beta_- + \boldsymbol{\alpha}_- \mathbf{p} + \frac{E^2 + \varepsilon_{\mathbf{p}}^2 - \varepsilon_{\mathbf{q}}^2}{2E}\right)\left(m\beta_+ + \boldsymbol{\alpha}_+ \mathbf{q} + \frac{E^2 - \varepsilon_{\mathbf{p}}^2 + \varepsilon_{\mathbf{q}}^2}{2E}\right)V(\mathbf{r}_1 - \mathbf{r}_2)\psi(\mathbf{r}_1, \mathbf{r}_2). \qquad (23)$$

Here $\hat{\mathbf{p}} = -i\nabla_{\mathbf{r}_1}$ and $\hat{\mathbf{q}} = -i\nabla_{\mathbf{r}_2}$ are the momentum of electron and positron, $\varepsilon_{\mathbf{p}}^2 = m^2 + \hat{\mathbf{p}}^2$ and $\varepsilon_{\mathbf{q}}^2 = m^2 + \hat{\mathbf{q}}^2$.

Eq. (23) is the bound-state equation for the electron-positron system with the symmetric representation of the particles in the quasi non-relativistic limit. Analysis for this equation is carried out in the next section.

## 5. The system at rest

In the rest of the system, $\hat{\mathbf{p}} = -\hat{\mathbf{q}} = -i\nabla_{\mathbf{r}}$ is the relative-motion momentum operator, and $\mathbf{r} = \mathbf{r}_1 - \mathbf{r}_2$ is the relative radius vector. As a result, equation (23) reduces to:

$$\left(E^2 - 4\hat{\varepsilon}_{\mathbf{p}}^2\right)\left(E - \frac{1}{2}V(\mathbf{r})\right)\psi(\mathbf{r}) = 2\left(m\beta_- + \boldsymbol{\alpha}_-\mathbf{p} + \frac{E}{2}\right)\left(m\beta_+ - \boldsymbol{\alpha}_+\mathbf{p} + \frac{E}{2}\right)V(\mathbf{r})\psi(\mathbf{r}). \quad (24)$$

For the states with $v/c \propto \alpha \ll 1$, the characteristic momenta of the particles are much smaller than their mass ($m \gg |\mathbf{p}|$). Then, from (24) we obtain the "unperturbed" equation:

$$\left(E^2 - 4\hat{\varepsilon}_{\mathbf{p}}^2\right)\left(E - \frac{1}{2}V(\mathbf{r})\right)\psi(\mathbf{r}) = 2\left(m\beta_- + \frac{E}{2}\right)\left(m\beta_+ + \frac{E}{2}\right)V(\mathbf{r})\psi. \quad (25)$$

The wave functions are sought in the forms:

$$\psi^{(1)} = \varphi^{(1)}\begin{pmatrix}w_-\\0\end{pmatrix}\begin{pmatrix}w_+\\0\end{pmatrix},\ \psi^{(2)} = \varphi^{(2)}\begin{pmatrix}0\\w_-\end{pmatrix}\begin{pmatrix}0\\w_+\end{pmatrix},\ \psi^{(3)} = \varphi^{(3)}\begin{pmatrix}w_-\\0\end{pmatrix}\begin{pmatrix}0\\w_+\end{pmatrix},\ \psi^{(4)} = \varphi^{(4)}\begin{pmatrix}0\\w_-\end{pmatrix}\begin{pmatrix}w_+\\0\end{pmatrix}$$

Here $w_\pm$ are the spinors. Substituting these functions in Eq. (25), we obtain:

$$\left(E^2 - 4m^2 - 4\hat{\mathbf{p}}^2\right)\left(E - \frac{1}{2}V(\mathbf{r})\right)\varphi^{(1)} = 2\left(m + \frac{E}{2}\right)^2 V(\mathbf{r})\varphi^{(1)}(\mathbf{r}), \quad (26)$$

$$\left(E^2 - 4m^2 - 4\hat{\mathbf{p}}^2\right)\left(E - \frac{1}{2}V(\mathbf{r})\right)\varphi^{(2)} = 2\left(m - \frac{E}{2}\right)^2 V(\mathbf{r})\varphi^{(2)}(\mathbf{r}), \quad (27)$$

$$\left(E^2 - 4m^2 - 4\hat{\mathbf{p}}^2\right)\left(E - \frac{1}{2}V(\mathbf{r})\right)\varphi^{(3),(4)} = -2\left(m^2 - \frac{E^2}{4}\right)V(\mathbf{r})\varphi^{(3),(4)}(\mathbf{r}), \quad (28)$$

First, we consider Eq. (26), which is reduced to the form:

$$\frac{-4\hat{\mathbf{p}}^2}{E(E+2m)}\left(E - \frac{1}{2}V(\mathbf{r})\right)\varphi^{(1)} = (2m - E + V(\mathbf{r}))\varphi^{(1)}(\mathbf{r}). \quad (29)$$

If in the left-hand side (29) we omit the potential $V(\mathbf{r})$ under the Laplacian and replace $E + 2m \to 4m$ then Eq. (29) is reduced to the Schrödinger equation for positronium:

$$\left(\frac{\hat{\mathbf{p}}^2}{m} + V(\mathbf{r})\right)\varphi^{(1)} = (E - 2m)\varphi^{(1)}(\mathbf{r}). \quad (30)$$

According to (30), the energies of the positronium states are:

$$E_n^{(1)} = 2m - \frac{me^4}{4\hbar^2 n^2}, \quad (31)$$

where $n$ is the principal quantum number. These states (32) lie just below the energy $2m$.

Thus, Eq. (24) includes the solutions corresponding to the well-known Ps states. Note that the potential energy in the left-hand side of Eq. (29), and the difference between $E$ and $2m$ lead to relativistic corrections. They should not be analyzed because a number of corrections have already been omitted in Eq. (29).

Here the following is important: in the standard theory of the electron-positron field the wave function of the kind $\psi^{(1)}(\mathbf{r}) = \varphi^{(1)} \begin{pmatrix} w_- \\ 0 \end{pmatrix} \begin{pmatrix} w_+ \\ 0 \end{pmatrix}$ is forbidden because in the non-relativistic limit ($|\mathbf{p}| \ll m$) the first two components, not the second two, vanish in the positron bispinors (see [3], page 383) or, in other words, the positrons are prohibited to scatter in the states of the upper continuum. As we have seen above, in the symmetric representation of the particles the function $\psi^{(1)}(\mathbf{r})$ represents exactly the states of positronium (31).

The function $\psi^{(2)}(\mathbf{r}) = \varphi^{(2)} \begin{pmatrix} 0 \\ w_- \end{pmatrix} \begin{pmatrix} 0 \\ w_+ \end{pmatrix}$ does not exist in the standard model because in the above limit the second two components, not the first two, vanish in the electron bispinors or, in other words, the electrons are prohibited to scatter in the states of the lower continuum. In the model of the symmetrical representation this function is possible. Eq. (27) for $\psi^{(2)}(\mathbf{r})$ is rewritten as:

$$\frac{-4\hat{\mathbf{p}}^2}{E(E-2m)}\left(E - \frac{1}{2}V(\mathbf{r})\right)\varphi^{(2)} = \left(-2m - E + V(\mathbf{r})\right)\varphi^{(2)}(\mathbf{r}). \tag{32}$$

Eq. (32) can be obtained from the Eq. (29) by the replacement $m \to -m$. However, properties of Eqs. (29) and (32) differ significantly. The potential energy under the Laplacian in the left-hand side of Eq. (29) leads to the $\delta$-functional feature that does not mean that there is a particularly strong interaction. The integral value of this term leads only to a relativistic correction. This term should be regarded as small in comparison with the Coulomb interaction. In Eq. (32) the potential energy under the Laplacian cannot be omitted, since at the negative energy it results in the singular point

$$r = -\frac{e^2}{E} \tag{33}$$

of the equation. The wave function is discontinuous at this singular point that leads to non-normalized wave function of the negative mass boson states.

Of course, Eq. (32) derived from the Bethe-Salpeter equation for a number of assumptions presented above. Beyond these assumptions, the existence of these negative mass states remains unclear.

Now explore Eq. (28). It is obviously that

$$\varphi^{(3)} = \varphi^{(4)}. \qquad (34)$$

Note that in the standard theory of the electron-positron field the state $\psi^{(3)}(\mathbf{r}) = \varphi^{(3)} \begin{pmatrix} w_- \\ 0 \end{pmatrix} \begin{pmatrix} 0 \\ w_+ \end{pmatrix}$ corresponds to the Ps one, and the state $\psi^{(4)}(\mathbf{r}) = \varphi^{(4)} \begin{pmatrix} 0 \\ w_- \end{pmatrix} \begin{pmatrix} w_+ \\ 0 \end{pmatrix}$ is prohibited. Eq. (28) is reduced to the form:

$$\frac{-4\hat{\mathbf{p}}^2}{(4m^2 - E^2)} \left( E - \frac{1}{2} V(\mathbf{r}) \right) \varphi^{(3)} = (E - V(\mathbf{r})) \varphi^{(3)}(\mathbf{r}). \qquad (35)$$

Due to the symmetry of the problem, the only dedicated energy is $E = 0$. Then Eq. (35) corresponds to a strongly coupled state, in which the mass of the composite boson vanishes completely because of the high binding energy, $-2m$, and the radial part of the wave function decreases exponentially with increasing modulus of the relative radius vector between the particles of the coupled pair:

$$\varphi^{(3)} \propto \sqrt{\lambdabar_e r} K_{l+1/2}(\sqrt{2} r / \lambdabar_e), \qquad (36)$$

where $\lambdabar_e$ is the Compton wavelength of the electron, $K_{l+1/2}$ is modified spherical Bessel function and $l = 0,1,2...$ is the orbital quantum number.

However, in this massless state the composite boson cannot be at rest, as is assumed in this section. Furthermore the assumptions used in the derivation of Eq. (23) cannot be applied for this state. In the next section we consider the interaction retardation and interaction through the vector potential, and study the massless state of the electron-positron system with the real coupling constant equal to $\alpha$.

## 6. The massless composite boson

We are searching for a solution $\psi(1,2)$ of Eq. (13) in the form of a stationary wave with the phase velocity equal to the speed of light. Let $\mathbf{p} + \mathbf{q} = \mathbf{g}$ and the momentum of the pair, $\mathbf{g}$, is directed along the z-axis. It is the strongly coupled state with the energy $E = g$. Due to the symmetry of the problem, we have to put $z_1 = z_2 = z$ and $t_1 = t_2 = t$ for this massless boson state that allows us to introduce the two-dimensional relative vector between the particles, $\boldsymbol{\rho} = \boldsymbol{\rho}_1 - \boldsymbol{\rho}_2$. Then the wave function can be represented as

$$\psi(1,2) = \varphi(\boldsymbol{\rho}, \mathbf{g}) \exp(igz - igt). \tag{37}$$

This wave function is an infinite line string, the transverse size of which is determined by $\varphi(\boldsymbol{\rho}, \mathbf{g})$.

Now we find the interaction function (9) for the boson state (37). The relativistic retarded function $\delta_+(s_{56}^2)$ considered in [9,10], takes in our case the form:

$$\delta_+(t_{56}^2 - \rho_{56}^2) = \frac{1}{4\pi\rho_{56}} \int_{-\infty}^{+\infty} \left( e^{-i\omega(t_5 - t_6)} + e^{-i\omega(t_6 - t_5)} \right) e^{i|\omega|\rho_{56}} d\omega, \tag{38}$$

where $\rho_{56} = |\boldsymbol{\rho}_5 - \boldsymbol{\rho}_6|$. It was taken into account that since the phase velocity of the wave (37) is equal to the velocity of light, for the stationary distribution of the charge density the interaction between the electron and positron can only occur in the same layers ($z_5 = z_6$), which are perpendicular to the wave vector $\mathbf{g}$ of (37). Considering that $\alpha_{-z}\alpha_{+z} = 1$ (here $\alpha_{-z}$ and $\alpha_{+z}$ are the $z$-component of the velocity operators for the electron and positron, respectively), the factor of $(1 - \boldsymbol{\alpha}_- \boldsymbol{\alpha}_+)$ in (9) should be replaced by $-\boldsymbol{\alpha}_{-\rho}\boldsymbol{\alpha}_{+\rho}$.

We can say that in this state the electron and positron not interact through the Coulomb potential and their retarded interaction occur through the vector potential which is due to the particles transverse motion defined by the function $\varphi(\boldsymbol{\rho}, \mathbf{g})$ in (37).

As a result, for the massless composite boson state (37) Eq. (13) is reduced to:

$$\varphi(\boldsymbol{\rho}_{12}, \mathbf{g}) e^{igz - igt} = -ie^2 \int_{-\infty}^{z} dz_3 \int_{-\infty}^{z} dz_4 \int d\boldsymbol{\rho}_3 \int d\boldsymbol{\rho}_4 \int_{-\infty}^{t} dt_3 \int_{-\infty}^{t} dt_4 \sum_{\mathbf{pq}} \frac{\exp(i\mathbf{p}(\mathbf{r}_1 - \mathbf{r}_3) + i\mathbf{q}(\mathbf{r}_2 - \mathbf{r}_4))}{4\varepsilon_p \varepsilon_q}$$

$$\int_{-\infty}^{+\infty} e^{i|\omega|\rho_{34}} d\omega \left\{ \Lambda_-^+(\mathbf{p}) e^{-i\varepsilon_p(t - t_3)} + \Lambda_-^-(\mathbf{p}) e^{i\varepsilon_p(t - t_3)} \right\} \left\{ \Lambda_+^+(\mathbf{q}) e^{-i\varepsilon_q(t - t_4)} + \Lambda_+^-(\mathbf{q}) e^{i\varepsilon_q(t - t_4)} \right\} \tag{39}$$

$$\left\{ e^{-i\omega(t_3 - t_4)} + e^{-i\omega(t_4 - t_3)} \right\} e^{i\frac{g}{2}(z_3 + z_4) - i\frac{g}{2}(t_3 + t_4)} \frac{1}{4\pi\rho_{34}} (\boldsymbol{\alpha}_{-\rho}\boldsymbol{\alpha}_{+\rho}) \varphi(\boldsymbol{\rho}_{34}, \mathbf{g})$$

Here $\Lambda_-^+(\mathbf{p}) = \varepsilon_\mathbf{p} + m\beta_- + \boldsymbol{\alpha}_- \mathbf{p}_\rho + p_z$ and $\Lambda_-^-(\mathbf{p}) = \varepsilon_\mathbf{p} - m\beta_- - \boldsymbol{\alpha}_- \mathbf{p}_\rho - p_z$ are the electron operators, $\Lambda_+^+(\mathbf{q}) = \varepsilon_\mathbf{q} + m\beta_+ + \boldsymbol{\alpha}_+ \mathbf{q}_\rho + q_z$ and $\Lambda_+^-(\mathbf{q}) = \varepsilon_\mathbf{q} - m\beta_+ - \boldsymbol{\alpha}_+ \mathbf{q}_\rho - q_z$ are the positron ones.

At first, analyzing only the $z$- dependent functions in (39), we integrate over $z_3$ and $z_4$:

$$e^{igz} = \sum_{p_z q_z} \int_{-\infty}^{z \to \infty} dz_3 \int_{-\infty}^{z \to \infty} dz_4 e^{i\frac{g}{2}(z_3+z_4)} \exp(ip_z(z-z_3)+iq_z(z-z_4))$$

$$= (2\pi)^2 \sum_{p_z q_z} \exp(i(p_z+q_z)z)\delta(p_z - \frac{g}{2})\delta(q_z - \frac{g}{2}) = e^{igz}\bigg|_{q_z=p_z=\frac{g}{2}}$$

Since the function $\varphi$ depends only on $\boldsymbol{\rho}_{34}$, in (39) we replace the integration variables: $\int d\boldsymbol{\rho}_3 \int d\boldsymbol{\rho}_4 = \int d\boldsymbol{\rho}_{34} \int d(\boldsymbol{\rho}_3+\boldsymbol{\rho}_4)/2$. Thereafter the integral over $(\boldsymbol{\rho}_3+\boldsymbol{\rho}_4)/2$ on the right side of (39) is easily calculated, and gives $(2\pi)^2\delta(\mathbf{p}_\rho+\mathbf{q}_\rho)$. Consequently Eq. (39) takes the form:

$$\varphi(\boldsymbol{\rho}_{12})e^{-igt} = -i\frac{e^2}{4\pi}\int\frac{d\boldsymbol{\rho}_{34}}{\rho_{34}}\int_{-\infty}^{t}dt_3\int_{-\infty}^{t}dt_4\sum_{\mathbf{p}_\rho}\frac{\exp(i\mathbf{p}_\rho(\mathbf{r}_{12}-\mathbf{r}_{34}))}{4\varepsilon_p^2}\int_{-\infty}^{+\infty}e^{i|\omega|\rho_{34}}d\omega$$

$$\left\{\Lambda_-^+(\mathbf{p})e^{-i\varepsilon_p(t-t_3)} + \Lambda_-^-(\mathbf{p})e^{i\varepsilon_p(t-t_3)}\right\}\left\{\Lambda_+^+(\mathbf{q})e^{-i\varepsilon_q(t-t_4)} + \Lambda_+^-(\mathbf{q})e^{i\varepsilon_q(t-t_4)}\right\} \quad (40)$$

$$\left\{e^{-i\omega(t_3-t_4)} + e^{-i\omega(t_4-t_3)}\right\}e^{-i\frac{g}{2}(t_3+t_4)}(\boldsymbol{\alpha}_{-\rho}\boldsymbol{\alpha}_{+\rho})\varphi(\boldsymbol{\rho}_{34})$$

where $\mathbf{q}_\rho = -\mathbf{p}_\rho$, $q_z = p_z = \frac{g}{2}$ and $\varepsilon_p = \varepsilon_q$.

Now integrating over $t_3$ and $t_4$, Eq. (40) is rewritten as:

$$\varphi(\boldsymbol{\rho}_{12}) = -i\frac{e^2}{4\pi}\int\frac{d\boldsymbol{\rho}_{34}}{\rho_{34}}\sum_{\mathbf{p}_\rho}\frac{\exp(i\mathbf{p}_\rho(\boldsymbol{\rho}_{12}-\boldsymbol{\rho}_{34}))}{4\varepsilon_p^2}\int_{-\infty}^{\infty}e^{i|\omega|\rho_{34}}d\omega I(\omega)(\boldsymbol{\alpha}_{-\rho}\boldsymbol{\alpha}_{+\rho})\varphi(\boldsymbol{\rho}_{34}), \quad (41)$$

where the function $I(\omega)$ has the three terms, $I(\omega) = I_1(\omega) + I_2(\omega) + I_3(\omega)$, which can be presented as:

$$I_1 = 2\frac{\Lambda_-^+(\mathbf{p})\Lambda_+^+(\mathbf{q})}{\omega^2 - (\varepsilon_p - \frac{g}{2})^2 + i\delta}, \quad I_2 = 2\frac{\Lambda_-^-(\mathbf{p})\Lambda_+^-(\mathbf{q})}{\omega^2 - (\varepsilon_q + \frac{g}{2})^2 - i\delta}$$

and

$$I_3 = \frac{\Lambda_-^-(\mathbf{p})\Lambda_+^+(\mathbf{q}) + \Lambda_-^+(\mathbf{p})\Lambda_+^-(\mathbf{q})}{-g} * \left(\frac{2\varepsilon_p - g}{\omega^2 - (\varepsilon_p - \frac{g}{2})^2 + i\delta} + \frac{-2\varepsilon_p - g}{\omega^2 - (\varepsilon_p + \frac{g}{2})^2 - i\delta}\right).$$

Here $\delta \to 0^+$ is the rule for bypassing simple poles.

All the three integrals over $\omega$ on the right side of (41) are easily calculated. In the end we get:

$$T_1 = \int_{-\infty}^{\infty} I_1(\omega) e^{i|\omega|\rho_{34}} d\omega = 8i \frac{\Lambda_-^+(\mathbf{p})\Lambda_+^+(\mathbf{q})}{2\varepsilon_p - g} \left(\cos(x)si(x) - \sin(x)ci(x)\right), \tag{42}$$

where $x = |\varepsilon_p - \frac{g}{2}|\rho_{34}$,

$$T_2 = \int_{-\infty}^{\infty} I_2(\omega) e^{i|\omega|\rho_{34}} d\omega = 8i \frac{\Lambda_-^-(\mathbf{p})\Lambda_+^-(\mathbf{q})}{2\varepsilon_p + g} \left\{\pi e^{iy} + \cos(y)si(y) - \sin(y)ci(y)\right\} \tag{43}$$

with $y = (\varepsilon_p + \frac{g}{2})\rho_{34}$ and

$$T_3 = \int_{-\infty}^{\infty} I_3(\omega) e^{i|\omega|\rho_{34}} d\omega = 4i \frac{\Lambda_-^-(\mathbf{p})\Lambda_+^+(\mathbf{q}) + \Lambda_-^+(\mathbf{p})\Lambda_+^-(\mathbf{q})}{g} *$$
$$\left\{\pi e^{iy} - \cos(x)si(x) + \sin(x)ci(x) + \cos(y)si(y) - \sin(y)ci(y)\right\} . \tag{44}$$

Here $si(x)$ and $ci(x)$ are the integral sine and cosine.

We now turn to the study of the bispinors of the function $\varphi(\boldsymbol{\rho}_{34})$ for the massless boson state given by (41). We need to find such spin functions on which the action of the following operators $\alpha_{-x}\alpha_{+x} + \alpha_{-y}\alpha_{+y}$, $\Lambda_-^+(\mathbf{p})\Lambda_+^+(\mathbf{q})$, $\Lambda_-^-(\mathbf{p})\Lambda_+^-(\mathbf{q})$ and $\Lambda_-^-(\mathbf{p})\Lambda_+^+(\mathbf{q}) + \Lambda_-^+(\mathbf{p})\Lambda_+^-(\mathbf{q})$ is reduced only to the multiplication of these functions on some scalars.

For the wave (37) the bispinors for each particle of the composite boson can be characterized by the projection of the particle spin on the momentum vector, or, in other words, the particle helicity. There are eight bispinors functions $\eta_i (i=1,...,8)$ for which the helicities of both the electron and positron are simultaneously either positive

$$\eta_{i=1,...,4} = \begin{pmatrix}1\\0\\0\\0\end{pmatrix}\begin{pmatrix}1\\0\\0\\0\end{pmatrix}, \begin{pmatrix}1\\0\\0\\0\end{pmatrix}\begin{pmatrix}0\\0\\1\\0\end{pmatrix}, \begin{pmatrix}0\\0\\1\\0\end{pmatrix}\begin{pmatrix}1\\0\\0\\0\end{pmatrix}, \begin{pmatrix}0\\0\\1\\0\end{pmatrix}\begin{pmatrix}0\\0\\1\\0\end{pmatrix},$$

or negative

$$\eta_{i=5,...,8} = \begin{pmatrix}0\\1\\0\\0\end{pmatrix}\begin{pmatrix}0\\1\\0\\0\end{pmatrix}, \begin{pmatrix}0\\1\\0\\0\end{pmatrix}\begin{pmatrix}0\\0\\0\\1\end{pmatrix}, \begin{pmatrix}0\\0\\0\\1\end{pmatrix}\begin{pmatrix}0\\1\\0\\0\end{pmatrix}, \begin{pmatrix}0\\0\\0\\1\end{pmatrix}\begin{pmatrix}0\\0\\0\\1\end{pmatrix}.$$

Because $\boldsymbol{\alpha}_{-\rho}\boldsymbol{\alpha}_{+\rho} = \alpha_{-x}\alpha_{+x} + \alpha_{-y}\alpha_{+y}$, one can convince that

$$\boldsymbol{\alpha}_{-\rho}\boldsymbol{\alpha}_{+\rho}\eta_i = 0.$$

This means that the interaction function $G^{(1)}(3,4;5,6) \propto \boldsymbol{\alpha}_{-\rho}\boldsymbol{\alpha}_{+\rho}$ vanishes for these states. That is, the massless boson state cannot be formed in the case when the helicities of these two particles are the same.

Because the functions $\psi^{(3),(4)}$ introduced above in Section 5, satisfy Eq. (28) with $E = 0$, by analogy with these functions we can present the required functions in the form:

$$\varphi_{1,2}(\mathbf{p}) = \chi_1(\mathbf{p})\left[\begin{pmatrix}1\\0\\0\\0\end{pmatrix}\begin{pmatrix}0\\0\\0\\1\end{pmatrix} + \begin{pmatrix}0\\0\\0\\1\end{pmatrix}\begin{pmatrix}1\\0\\0\\0\end{pmatrix}\right], \quad \chi_2(\mathbf{p})\left[\begin{pmatrix}0\\1\\0\\0\end{pmatrix}\begin{pmatrix}0\\0\\1\\0\end{pmatrix} + \begin{pmatrix}0\\0\\1\\0\end{pmatrix}\begin{pmatrix}0\\1\\0\\0\end{pmatrix}\right]. \tag{45}$$

Here $\chi_{1,2}(\boldsymbol{\rho},\mathbf{g})$ are the coordinate wave functions of the transverse motion of the strongly coupled electron-positron pair.

In the states (45) the helicities of the electron and positron are opposite, and if the electron state is defined by the top spinor then the positron state is determined by the bottom spinor and vice versa. For the functions (45) we have:

$$\boldsymbol{\alpha}_{-\rho}\boldsymbol{\alpha}_{+\rho}\varphi_{1,2} = 2\varphi_{1,2},$$

$$\Lambda_-^+(\mathbf{p})\Lambda_+^+(\mathbf{q})\varphi_{1,2} = g(\varepsilon_\mathbf{p} + \frac{g}{2})\varphi_{1,2},$$

$$\Lambda_-^-(\mathbf{p})\Lambda_+^-(\mathbf{q})\varphi_{1,2} = g(-\varepsilon_\mathbf{p} + \frac{g}{2})\varphi_{1,2},$$

$$(\Lambda_-^-(\mathbf{p})\Lambda_+^+(\mathbf{q}) + \Lambda_-^+(\mathbf{p})\Lambda_+^-(\mathbf{q}))\varphi_{1,2} = (4\varepsilon_\mathbf{p}^2 - g^2)\varphi_{1,2}.$$

As a result, Eq. (41) is transformed to the following integral equation on the coordinate function $\chi_{1,2}$ (since $\chi_1 = \chi_2$ the lower index 1,2 can be omitted):

$$\chi(\boldsymbol{\rho}_{12}) = \frac{e^2}{(2\pi)^3}\int\frac{d\boldsymbol{\rho}_{34}}{\rho_{34}}\int\frac{d\mathbf{p}_\rho}{\varepsilon_p^2}\exp(i\mathbf{p}_\rho(\boldsymbol{\rho}_{12} - \boldsymbol{\rho}_{34}))(T_1 + T_2 + T_3)\chi(\boldsymbol{\rho}_{34}), \tag{46}$$

where

$$T_1 = g\frac{2\varepsilon_\mathbf{p} + g}{2\varepsilon_p - g}(\cos(x)si(x) - \sin(x)ci(x)), \tag{47}$$

$$T_2 = g \frac{g - 2\varepsilon_\mathbf{p}}{g + 2\varepsilon_p} \left\{ \pi e^{iy} + \cos(y)si(y) - \sin(y)ci(y) \right\} \tag{48}$$

and

$$T_3 = \frac{4\varepsilon_\mathbf{p}^2 - g^2}{g} \left\{ \pi e^{iy} - \cos(x)si(x) + \sin(x)ci(x) + \cos(y)si(y) - \sin(y)ci(y) \right\}. \tag{49}$$

Eq. (46) with the notations (47)-(49) is a rather complicated integral equation for the wave function $\chi(\boldsymbol{\rho}_{12})$ of the transverse motion of the bound pair. It seems important to show that this equation has solutions for the normalized eigenfunctions $\chi(\boldsymbol{\rho}_{12}, g)$. Below we will demonstrate it for the case of the *S*-state of the bound pair and small momenta of the boson, $g \ll m$.

It should be noted that just this case is important for experimental discovery of this massless boson in the well-known physical process, which will be discussed below.

For the *S*-state the angular momentum of the relative motion of the bound pair is zero, and the function $\chi(\boldsymbol{\rho}_{12}, g)$ depends only on the modulus of the relative vector, that is $\chi(\rho_{12}, g)$. Then after integrations over the azimuthal angle of the vector $\boldsymbol{\rho}_{34}$ and over the azimuthal angle of the vector $\mathbf{p}_\rho$ Eq. (46) is reduced to:

$$\chi(\rho_{12}) = \frac{e^2}{2\pi} \int_0^\infty d\rho_{34} \int_0^\infty p_\rho dp_\rho J_0(p_\rho \rho_{12}) J_0(p_\rho \rho_{34}) \frac{\sum_{i=1}^3 T_i(q, \rho_{34})}{\varepsilon_q^2} \chi(\rho_{34}). \tag{50}$$

In the case $g \ll m$ from (47)-(49) we find that $T_3 \gg T_1, T_2$ and

$$T_3 \cong 4\pi \frac{\varepsilon_q^2}{g} \exp(i\varepsilon_q \rho_{34}). \tag{51}$$

Substituting (51) to Eq. (50), the latter equation is reduced to the homogeneous Fredholm integral equation of the second kind:

$$\chi(\rho_{12}) = 2\frac{e^2}{g} \int_0^\infty d\rho_{34} R(\rho_{12}, \rho_{34}) \chi(\rho_{34}) \tag{52}$$

with the kernel

$$R(\rho_{12}, \rho_{34}) = \int_0^\infty p_\rho dp_\rho J_0(p_\rho \rho_{12}) J_0(p_\rho \rho_{34}) \exp(i\varepsilon_p \rho_{34}). \tag{53}$$

where $J_0$ is the Bessel function of the first kind.

It can be seen that the kernel (53) is non-Fredholm one, and according to the asymptotic property of the Bessel function $J_0$, can only be defined as the principal value integral. That is the equation (52) with the kernel (53) is still difficult to study.

The easier situation occurs when the Fourier transform of the $\chi$-functions is used, $\chi(\mathbf{q}) = \int \chi(\boldsymbol{\rho}_{12})\exp(-i\mathbf{q}\boldsymbol{\rho}_{12})d\boldsymbol{\rho}_{12}$. Then from (46) we obtain:

$$\chi(\mathbf{q}) = \frac{e^2}{(2\pi)^3}\int \frac{d\boldsymbol{\rho}_{34}}{\rho_{34}} \frac{\sum_{i=1}^{3} T_i(q,\rho_{34})}{\varepsilon_q^2} \int d\mathbf{f}\, e^{i(\mathbf{f}-\mathbf{q})\boldsymbol{\rho}_{34}} \chi(\mathbf{f}). \tag{54}$$

For the $S$-state of the bound pair Eq. (54) can be written as:

$$\chi(q) = \frac{e^2}{2\pi}\int d\rho_{34} \frac{\sum_{i=1}^{3} T_i(q,\rho_{34})}{\varepsilon_q^2} J_0(q\rho_{34})\int f J_0(f\rho_{34})\chi(f)df. \tag{55}$$

In the case $g \ll m$ we can use the function (51). Then Eq. (55) is reduced to the homogeneous Fredholm integral equation of the second kind:

$$\chi(q) = \frac{e^2}{g}\int_0^\infty Q(q,f)\chi(f)df \tag{56}$$

with the kernel

$$Q(q,f) = 2f\int_0^\infty J_0(q\rho_{34})J_0(f\rho_{34})\exp(i\varepsilon_q \rho_{34})d\rho_{34}. \tag{57}$$

Now the integral in the right-hand side of (57) that has a relationship with the discontinuous Weber-Schafheitlin integral, is absolutely convergent integral which, however, is expressed through a discontinuous function, as will see below.

It is convenient to use the dimensionless variables: $x \to \rho_{34}/\lambdabar_e$, $q \to q\lambdabar_e$, $f \to f\lambdabar_e$ and $g \to gm$. Then Eqs. (56)-(57) are rewritten as:

$$\chi(q) = \frac{\alpha}{g}\int_0^\infty Q(q,f)\chi(f)df \tag{58}$$

and

$$Q(q,f) = 2f\int_0^\infty J_0(qx)J_0(fx)\exp(ix\sqrt{1+q^2})dx. \tag{59}$$

Here $\alpha$ is the fine structure constant.

The integral on the right side of (59) was previously calculated [28], and thus, the kernel $Q(q,f) = \operatorname{Re} Q + i \operatorname{Im} Q$ is:

$$\operatorname{Im} Q = \begin{cases} \dfrac{4f}{\pi\sqrt{1-f^2+2fq}} \mathbf{K}\!\left(\dfrac{2\sqrt{fq}}{\sqrt{1-f^2+2fq}}\right), & f < \sqrt{1+q^2}-q \\[1em] \dfrac{2}{\pi}\sqrt{\dfrac{f}{q}} \mathbf{K}\!\left(\dfrac{\sqrt{1-f^2+2fq}}{2\sqrt{fq}}\right), & \sqrt{1+q^2}-q < f < \sqrt{1+q^2}+q \\[1em] 0, & f > \sqrt{1+q^2}+q \end{cases} \quad (60)$$

and

$$\operatorname{Re} Q = \begin{cases} 0, & f < \sqrt{1+q^2}-q \\[1em] \dfrac{2}{\pi}\sqrt{\dfrac{f}{q}} \mathbf{K}\!\left(\dfrac{\sqrt{f^2+2fq-1}}{2\sqrt{fq}}\right), & \sqrt{1+q^2}-q < f < \sqrt{1+q^2}+q \\[1em] \dfrac{4f}{\pi\sqrt{f^2+2fq-1}} \mathbf{K}\!\left(\dfrac{2\sqrt{fq}}{\sqrt{f^2+2fq-1}}\right), & f > \sqrt{1+q^2}+q \end{cases} \quad (61)$$

Here $\mathbf{K}$ is the complete elliptic integral of the first kind. Since this kernel, having the weak logarithmic singularity, is complex, the boson wave function $\chi(q)$ is complex also. Partition of $\chi(q)$ into the imaginary and real parts is in a sense arbitrary because the wave function satisfies the phase transformation $\chi(q) \to \chi(q)e^{i\phi}$.

It is easy to see that the kernel (60)-(61) is not the Fredholm kernel. Therefore we can expect that the spectrum of the characteristic numbers of the non-Fredholm kernel (60)-(61) can be continuous, that is, the characteristic numbers can occupy the whole intervals $\{g\}$. Accordingly, these intervals can be corresponded to the continuous spectra of the eigenfunctions $\chi(q,g)$.

To study the composite massless boson states, below we use the following procedure. The complex eigenfunctions should be normalized, $2\pi \int_0^\infty |\chi(q,g)|^2 qdq = 1$. Hence, $\chi(q \to \infty, g) \to 0$. Assume that the wave function goes to zero fast enough so that Eq. (58) can be replaced as:

$$\chi(q) = \frac{\alpha}{g} \int_0^{f_0} Q(q,f)\chi(f)df, \qquad (62)$$

where the value of $f_0(g)$ in units of $\lambda_e^{-1}$ depends on the boson kinetic energy, and $q \in [0, f_0]$.

The kernel $Q(q,f)$ given by (60)-(61), is replaced by the square $N \times N$ matrix:

$$Q_{ij}^{(r)} + iQ_{ij}^{(i)} = \frac{f_0}{N-1} Q(q_i, f_j),$$

which is denoted as $\mathbf{Q}^{(r)} + i\mathbf{Q}^{(i)}$. Here $N$ is the partition number of the interval $[0, f_0]$. The function $\chi(q)$ is replaced by the two $N$-dimensional vectors $\chi(q) \Rightarrow \chi^{(r)} + i\chi^{(i)}$. Then Eq. (62) is reduced to two related linear equations:

$$\left(\frac{g}{\alpha}\mathbf{I} - \mathbf{Q}^{(r)}\right)\chi^{(r)} = -\mathbf{Q}^{(i)}\chi^{(i)}, \qquad (63)$$

$$\left(\frac{g}{\alpha}\mathbf{I} - \mathbf{Q}^{(r)}\right)\chi^{(i)} = \mathbf{Q}^{(i)}\chi^{(r)}. \qquad (64)$$

Here $\mathbf{I}$ is the unit $N \times N$ matrix.

From (63)-(64) we obtain the homogeneous system of linear $N$-equations for the vector $\chi^{(r)}$:

$$\mathbf{A}\chi^{(r)} = 0, \qquad (65)$$

where the $\mathbf{A}$ matrix is:

$$\mathbf{A} = \frac{g}{\alpha}\mathbf{I} - \mathbf{Q}^{(r)} + \mathbf{Q}^{(i)}\left(\frac{g}{\alpha}\mathbf{I} - \mathbf{Q}^{(r)}\right)^{-1}\mathbf{Q}^{(i)} \qquad (66)$$

Here $\left(\frac{g}{\alpha}\mathbf{I} - \mathbf{Q}^{(r)}\right)^{-1}$ means the inverse of $\left(\frac{g}{\alpha}\mathbf{I} - \mathbf{Q}^{(r)}\right)$.

In contrast to the Fredholm procedure for the integral equation with the Fredholm kernel, Eq. (65) is now required to introduce a boundary condition. It is sufficient to put

$$\mathrm{Re}\,\chi(f_0) = \delta. \qquad (67)$$

Here $\delta$ is a small value. Typically this value is assumed to be equal to $\delta = 10^{-6}$. This value does not matter because of the subsequent normalization of the wave function.

Using the boundary condition (67), from Eq. (65) we find the vector $\chi^{(r)}$ which is the real part of the boson wave function. Then, from the equation (64) reduced to the form:

$$\chi^{(i)} = \left(\frac{g}{\alpha}\mathbf{I} - \mathbf{Q}^{(r)}\right)^{-1} \mathbf{Q}^{(i)} \chi^{(r)}, \qquad (68)$$

the vector $\chi^{(i)}$ representing the imaginary part of the massless boson wave function, is determined. After normalization of the wave function $\chi^{(r)} + i\chi^{(i)}$ that is served as the initial approximation, the equation (63) is represented as

$$\chi^{(r)} = -\left(\frac{g}{\alpha}\mathbf{I} - \mathbf{Q}^{(r)}\right)^{-1} \mathbf{Q}^{(i)} \chi^{(i)}, \qquad (69)$$

and the system of two interrelated equations (68) and (69) was solved by the iterative method. The number of iterations was of the order of a few hundred to provide the convergent solution.

Then we found the average transverse momentum of the transverse motion of the strongly coupled electron-positron pair:

$$q_{av}\lambda_e = 2\pi \int_0^\infty |\chi(q,g)|^2 \, q^2 dq. \qquad (70)$$

Finally, having the wave function in the momentum representation, we obtained the wave function in the coordinate representation:

$$\chi(x) = \int_0^\infty q J_0(xq) \chi(q) dq \qquad (71)$$

and the average transverse radius of the massless boson wave function ($x = \rho_{34}/\lambda_e$):

$$\rho_{av}/\lambda_e = 2\pi \int_0^\infty |\chi(x,g)|^2 \, x^2 dx. \qquad (72)$$

Substituting the normalized function $\chi(q)$ into (71), the wave function $\chi(x)$ obtained was always normalized.

For all results presented below, $N = 3501$ was used. To obtain reproducible results, the upper limit of integration $f_0$ (in units of $\lambda_e^{-1}$) has a bottom restriction which depends on the kinetic energy of the boson. We used the relationship $f_0(g) = 84g + 50$ where $g$ is the dimensionless kinetic energy of the boson.

Fig. 1 shows the main result: the average transverse momentum $q_{av}$ and the average transverse radius $\rho_{av}$ as functions of the massless boson kinetic energy. In the low energy region, $149\,\mathrm{eV} \leq g \leq 400\,\mathrm{eV}$, the average transverse momentum $q_{av}$ approaches monotonically to $\lambdabar_e^{-1}$ with decreasing the energy, and the average transverse radius obeys logarithmic dependence, $\rho_{av} \propto -\log(g)$. Apparently this behavior occurs at $g \to 0$ that is consistent with the fact that the massless particles cannot be at rest.

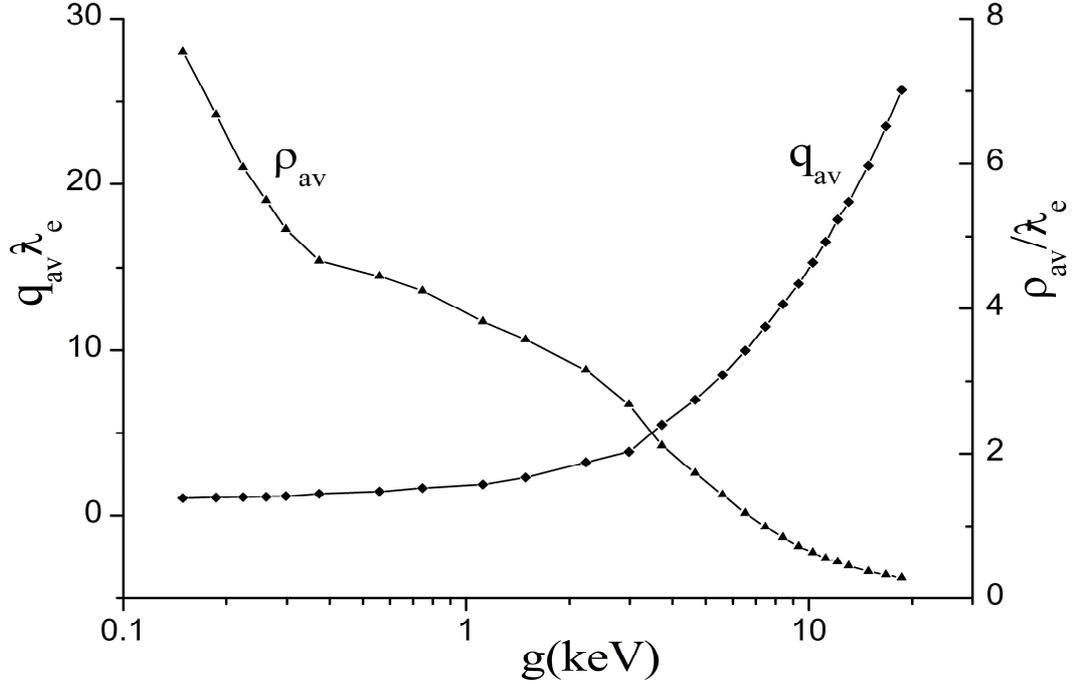

Fig. 1. The average transverse momentum $q_{av}$ (70) and the average transverse radius $\rho_{av}$ (72) as functions of the massless boson kinetic energy.

Features of the massless boson states in this energy region are demonstrated in Fig. 2. For the kinetic energy $g = 224\,\mathrm{eV}$ (the dimensionless value of $g = 0.06\alpha$) the wave function in the momentum space, $\chi(q) = \chi^{(r)} + i\chi^{(i)}$, is shown in Fig. 2a, and the wave function in the $\rho$ – coordinate space, $\chi(\rho,\mathbf{g})$, is presented in Fig.2b. Above we noted the arbitrary choice of the imaginary and real parts of $\chi$. Therefore we do not introduce the corresponding notations for the curves shown. One of these components is small as compared with the second, and this feature is always manifested for the wave functions in the low kinetic energy region. The effective transverse momentum found is equal to $q_{av} = 1.10\lambdabar_e^{-1}$ that is close to the inverse value of the Compton wave length of the electron. The sawtooth-like shape of the wave function near $q\lambdabar_e \approx 0.3$ correlates with the finite step $\Delta f = f_0/(N-1) \cong 0.0149$. After the transformation (71) the sawtooth-like features vanish. The effective transverse radius found is equal to $\rho_{av} = 5.95\lambdabar_e$.

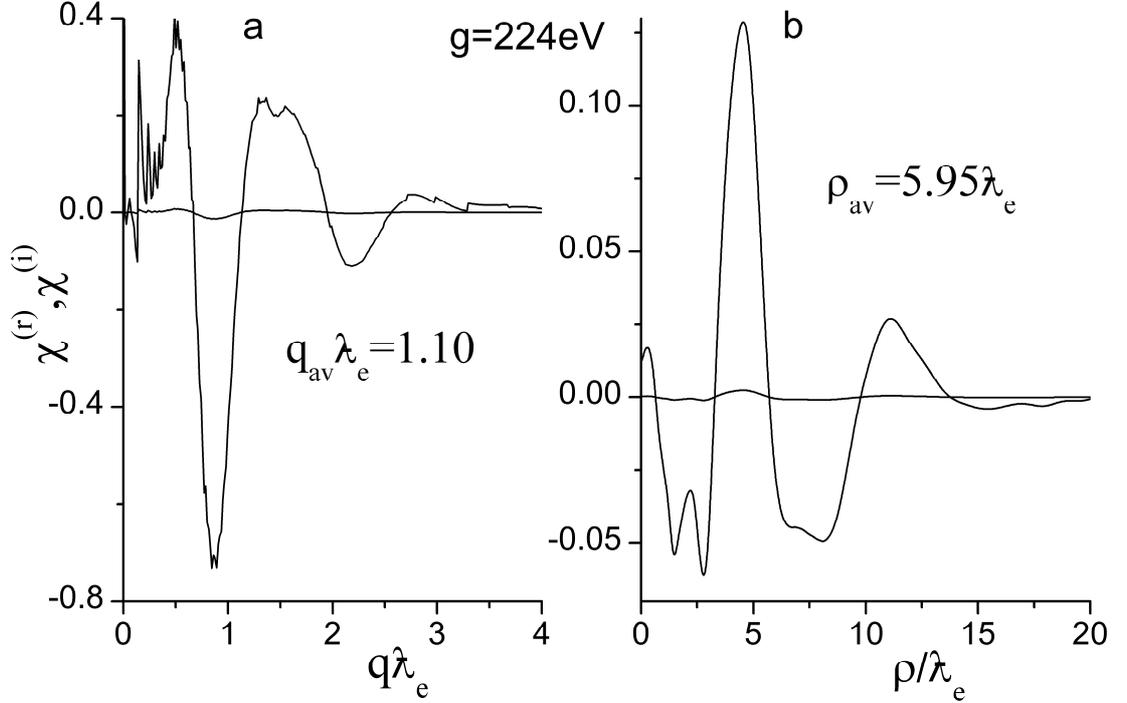

Fig. 2. The momentum space wave function (a) and the coordinate space wave function (b) for the massless boson kinetic energy equal to 224 eV.

With decreasing the energy $g < 224$ eV this sawtooth-like shape becomes more pronounced in the momentum space. In this case $q_{av}\lambdabar_e$ remains very close to 1, and $\rho_{av}$ increases more strongly as compared with the logarithmic dependence. However it may be only due to the limitation of the number $N$.

With increasing the kinetic energy the average momentum of the transverse motion of the strongly coupled electron-positron pair increases, and the value of $\rho_{av}$ characterized the localization of the wave function in the $\rho$ – coordinate space, decreases, as shown in Fig. 1. It means that the particles become closer to each other in this coordinate space.

In the middle region in Fig. 1, $0.4\,\text{keV} < g < 2\,\text{keV}$, both these values, $q_{av}$ and $\rho_{av}$, vary smoothly. At the same time the real and imaginary parts of the wave function become comparable.

This change in the wave functions clearly demonstrates Fig. 3, the data of which correspond to the energy $g = 1.12\,\text{keV}$ (the dimensionless value of $g = 0.3\alpha$). While at $g = 0.5\,\text{keV}$ one component of the wave function is approximately 3 times smaller than the other, for $g = 1.12\,\text{keV}$ both the components are of the same order. Comparing the data in Fig. 2 and Fig. 3, one can conclude that with increasing the kinetic energy oscillations of the wave functions in both the momentum and $\rho$ – coordinate spaces are enhanced.

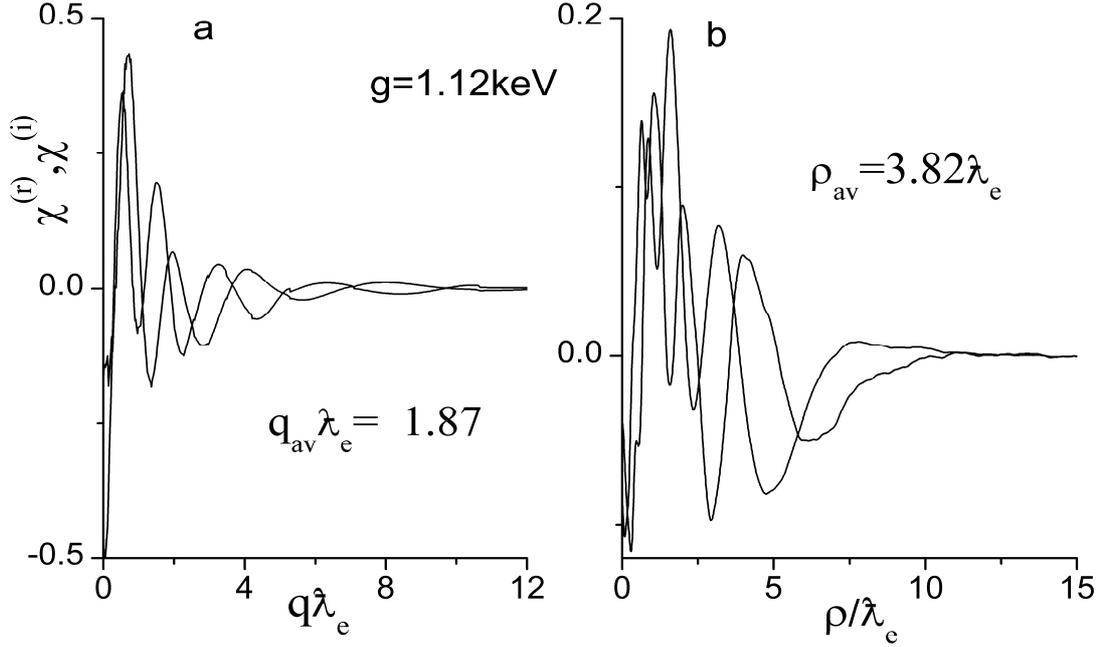

Fig. 3. The same as in Fig. 2 except for the massless boson kinetic energy equal to 1.12 keV.

As shown in Fig. 1, starting with energy $g \approx 4\,\text{keV}$ the effective transverse momentum of the bound electron-positron pair is sharply increased, and the spatial contraction of the wave function $\chi(\rho,g)$ becomes more pronounced. Fig. 4 shows the wave functions in both the momentum and coordinate spaces for the kinetic energy $g = 7.46\,\text{keV}$ (the dimensionless value of $g = 2\alpha$). The wave function in the momentum space presented in Fig. 4a, is more extended as compared with that for $g = 1.12\,\text{keV}$ (see Fig. 3a). Consequently, the effective transverse momentum is increased with the value $q_{av} = 1.87\lambdabar_e^{-1}$ for $g = 1.12\,\text{keV}$ to $q_{av} = 11.41\lambdabar_e^{-1}$ at $g = 7.46\,\text{keV}$. The spatial contraction of the wave function is very evident when comparing the data shown in Fig. 3b and Fig. 4b, according to which the transverse size of the wave function has decreased from $\rho_{av} = 3.82\lambdabar_e$ to the value $\rho_{av} = 0.99\lambdabar_e$. Note the additional phase factor $e^{i\pi}$ in the wave function which appeared by itself in our calculations.

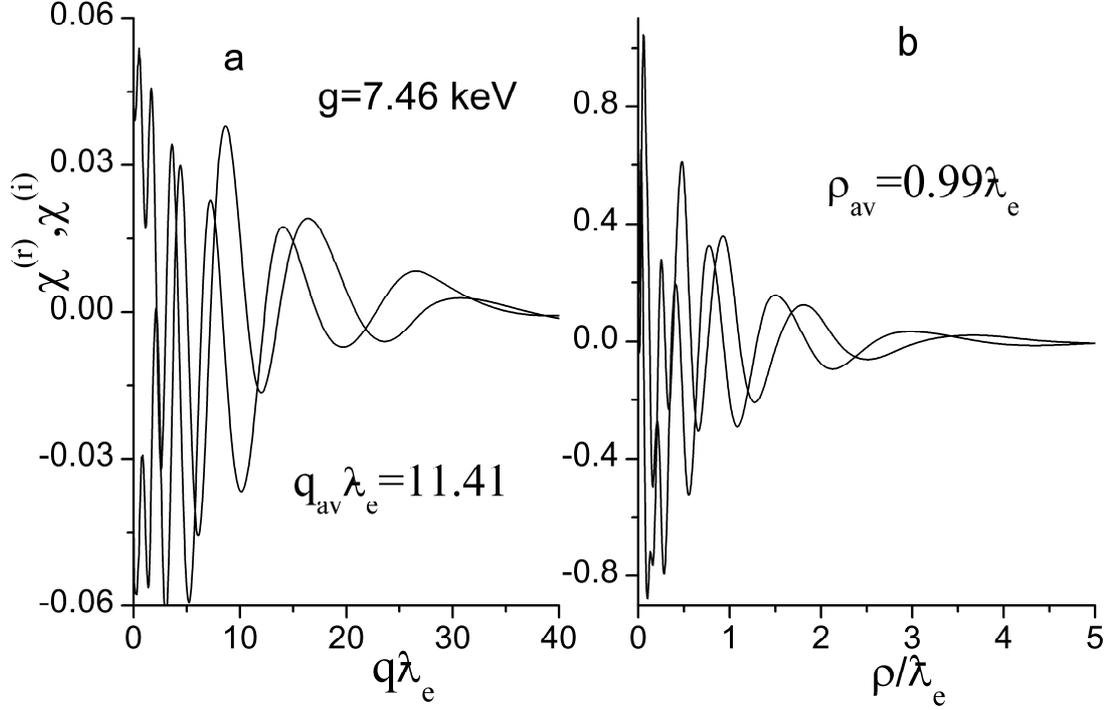

Fig. 4. The same as in Fig. 2 except for the massless boson kinetic energy equal to 7.46 keV.

Because $f_0$ increases linearly with $g$, and the partition number $N = 3501$ is fixed (with increasing $N$ calculation times increase very sharply), the massless boson kinetic energy, for which we could find the wave function, is restricted. However, with increasing energy there is another feature of the change of the wave function. Fig. 5 shows the boson state, corresponding to the kinetic energy $g = 18.64$ keV (the dimensionless value of $g = 5\alpha$). Comparing Fig. 5a with Fig. 4a, one can see that the extension of the wave function increases in the momentum space, the oscillation amplitudes decrease, and the wave function is grouped near the relatively small momenta. The average transverse momentum is equal to $q_{av} = 25.67\lambdabar_e^{-1}$ which is significantly more than that for $g = 7.46$ keV. Note that for small $q$ discrete steps on the curves are beginning to appear that is associated with the relatively large value $f_0 = 470$. The wave function in coordinate space is presented in Fig. 5b. By comparison with the data in Fig. 4b we should note the sharp contraction of the wave function in the coordinate space, and its localization in the region of the smaller distances between the electron and positron. So, the average transverse radius of the massless boson wave function is equal to $\rho_{av} = 0.29 \lambdabar_e$.

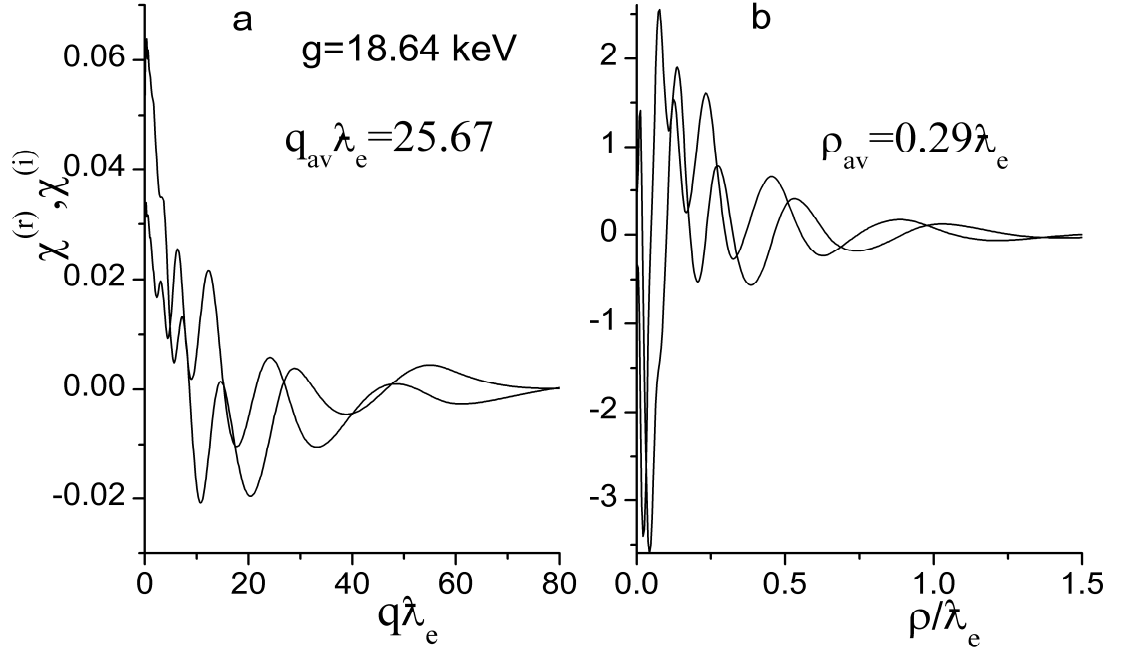

Fig. 5. The massless boson wave function in the momentum space (a) and the coordinate space (b) for the kinetic energy 18.64 keV.

## 7. Discussion and Conclusion

Assuming that the electron and positron are independent, different particles, each of these particles was characterized by the complete set of the Dirac plane waves. This leads inevitably to another choice of the free particle propagator with compared to that currently being used in QED. Then the bound-state Bethe-Salpeter equation in the ladder approximation with these free propagators was investigated.

In this symmetric representation one could expect states, which have a certain symmetry relative to Ps states, but have negative masses. For these states the equal-time bound-state equation was derived neglecting the interaction retardation and interaction through the vector potential. It turned out that the wave functions of the negative mass boson states are not normalized. Beyond these assumptions, the existence of these negative mass states remains unclear.

The branch of the massless composite bosons formed by the bound electron–positron system with the actual coupling equal to the fine structure constant states, have been found. Summarizing the results presented in Fig. 1-5 for the case of the *S*-state of the coupled electron-positron pair and small momenta of the composite boson, $g \ll m$, the following conclusions can be made:

1) The massless boson wave functions are complex-valued and normalized;

2) The average distance between the electron and positron diverges as the boson kinetic energy goes to zero that explains that the massless compound boson cannot be at rest;

3) The spatial contraction of the wave function of the transverse motion of the coupled electron-positron pair is continuously occurred with increasing the boson kinetic energy.

The existence of the massless bosons should lead to a process that is similar to the known process of the electron and positron annihilation, but have a significant distinguishing feature. Consider two colliding electron and positron beams with zero total momentum of the electron and positron from the beams. For the annihilation process nothing remains from the electron and positron, and two gamma quanta generated are emitted in opposite directions (the zero-spin channel is considered for clarity). The similar process involving the massless boson, denoted as $(e^- e^+)_{mb}$, is as follows:

$$e^- + e^+ \to (e^- e^+)_{mb} + \gamma_1 + \gamma_2. \qquad (45)$$

In contrast, the reaction should be characterized by the finite angular and momentum correlations of the radiation spectra that is due to the massless boson momentum. The total energy of the generated two gamma quanta and boson should be approximately equal to $2mc^2$. The same process can occur when the initial state is given by the para-positronium states.

At present, intense monoenergetic beams of positrons have been built in many laboratories (see [27-30] and references therein). The beam with the positron energy of 200 eV was used in [27] to obtain measurements of the two-dimensional angular correlation of the $2\gamma$ annihilation radiation from a clean Al(100) surface. The data shown in Fig. 3 of [27], presented the bell-shaped curves with the correlation angle up to 10 mrad. It was noted that the results obtained did not support the prevailing models of positrons annihilating on metal surfaces. From the standpoint of the reaction (45) one can assume that this experimental setup was made it possible to register this process with generation of the massless bosons up to the energy 5.11 keV.

Now this is only the assumption. If experiments carried out under vacuum with two colliding low-energy beams of electrons and positrons with equal energies, will show that the annihilation angular correlation spectra observed have finite widths (about 10 mrad, there is the reason for this) then this would lead to the discovery of the massless boson formed by the strongly coupled electron-positron pair.

**References**

1. Akhiezer, A. I. and Berestetskii, V. B. *Quantum Electrodynamics*. New York: Interscience Publishers, 1965.
2. Levich, V.G., Vdovin, Yu.A., and Myamlin, V.A., *Course of Theoretical Physics*, vol. 2, Moscow: Nauka, 1971, p. 473 (in Russian).
3. Berestesky V.B., Lifshits E.M., Pitaevsky L.P. *Quantum electrodynamics*, Fizmatlit, Moscow, 2002 (in Russian).
4. Feynman R.P. and Weinberg S. *Elementary particles and the laws of physics*. Cambridge University Press, 1999.
5. Feynman R.P. Phys. Rev. **76**, 749 (1949); Phys. Rev. **76**, 769 (1949).
6. Dirac P.A.M. *The principles of quantum mechanics*. Oxford, The Clarendon Press, 1958.
7. Zel'dovich Ya.B., Sov. Phys. Usp. **24,** 216 (1981).
8. Zel'dovich Ya.B., Popov V.S., Sov. Phys. Usp. **14,** 673 (1972).
9. Migdal A.B., Sov. Phys. Usp. **20,** 879 (1977).
10. Tamm I.E., Z. Phys. 62, 545 (1930).
11. Lukes T., Roberts M., Phys. Stat. Sol. **35,** 397 (1969).
12. Keldysh L.V., JETF **18,** 253 (1964).
13. Pantelides S.T., Rev. Mod. Phys. **50,** 797 (1978).
14. Salpeter E.E. and Bethe H.A., Phys. Rev. **84**, 1232 (1951).
15. Gell-Mann M. and Low P., Phys. Rev. **84**, 350 (1951).
16. Goldstein J.S., Phys. Rev. **91**, 1516 (1953).
17. Nakanishi N., Supp. Prog. Theor. Phys. No 43, 1 (1969).
18. Seto N., Prog. Theor. Phys. Suppl. No 95, 25 (1988).
19. Nishimura N., Higashijima K., Prog. Theor. Phys. **56**, 908 (1976).
20. Suttorp L. G., Ann. of Phys. **113**, 257 (1978).
21. Lucha W., Rupprecht H. and Schiiberl F., Phys. Rev. D**44**, 242 (1991).

22. Dae Sung Hwang, Karmanov V.A., Nucl. Phys. B **696**, 413 (2004).

23. Efimov G.V., Few-Body Systems, Volume 33, Issue 4, pp. 199-217 (2003).

24. Bawin M., Cugnon J., Sazdjian H., Int. J. Mod. Phys. A **11**, 5303 (1996).

25. Sazdjian H., Int. J. Mod. Phys. A **03**, 1235 (1988).

26. W. Greiner, J. Reinhardt. *Quantum Electrodynamics*. Springer-Verlag, Berlin Heidelberg 2009, p. 329.

27. D.D. Ivanenko. The preface to the book: *The latest development of quantum electrodynamics*. Foreign Literature Publishing House, Moscow, 1954, p. III (in Russian).

28. A. P. Prudnikov, Y. A. Brychkov and O. I. Marichev, *Integrals and Series*, *Special Functions,* Vol. 2 (Fizmatlit, Moscow, 2003), p. 199 (in Russian).

29. K. G. Lynn, A. P. Mills, Jr., R. N. West, S. Berko, K. F. Canter, and L. O. Roellig. Phys. Rev. Lett. 54, 1702 (1985).

30. P. Coleman. *Positron beams and their applications*. Singapore, World Scientific, 1999, 322 p.

31. M. Charlton. *Perspectives on Physics with Low Energy Positrons: Fundamentals, Beams and Scattering*, in: New Directions in Antimatter Chemistry and Physics, eds. C.M. Surko and F.A. Gianturco, Springer Netherlands, Kluwer Academic Publishers 2001.